\documentclass{article}
\usepackage{amsmath}
\usepackage{amsfonts}

\advance \textheight by \topskip
\advance \textheight by 1pt
\makeatother
\def\bea{\begin{eqnarray}}
\def\ena{\end{eqnarray}}
\def\non{\nonumber}

\def\lar{\longrightarrow}

\def\deg{\hbox{deg}}

\def\Ker{\hbox{Ker}}

\def\ch{\hbox{ch}}
\def\sgn{\hbox{sgn}}

\def\lar{\longrightarrow}
\def\tpsi{\tilde{\psi}}
\def\tH{\tilde{H}}
\def\e{\text{e}}

\def\uqs1{U_{\sqrt{-1}}(\widehat{sl_2})}
\def\upqs1{U'_{\sqrt{-1}}(\widehat{sl_2})}
\def\vis1{V_{\sqrt{-1}}(\Lambda_i)}

\def\bS{\bar{S}}
\def\ev{\text{ev}}
\def\vol{\text{vol}}
\newcommand{\qed}{\hbox{\rule{6pt}{6pt}}}
\newtheorem{prop}{Proposition}
\newtheorem{theorem}{Theorem}

\newtheorem{lemma}{Lemma}
\newtheorem{cor}{Corollary}
\newtheorem{conj}{Conjecture}

\title{
On the Space of KdV Fields
}

\author{
Atsushi Nakayashiki\thanks{
Department of Mathematics,
Kyushu University,
Ropponmatsu 4-2-1, Fukuoka 810-8560, Japan, \quad 
e-mail: 6vertex@math.kyushu-u.ac.jp, Mathematics Subject Classifications: 37K20,17B69,17B37, Key words: KdV hierarchy, boson-fermion correspondence, 
space of fields, D-module.}
}

\date{}
\begin{document}
\maketitle
\begin{abstract}
The space of functions $A$ over the phase space of KdV-hierarchy
is studied as a module over the ring ${\cal D}$ generated by
commuting derivations. A ${\cal D}$-free resolution of $A$ 
is constructed by Babelon, Bernard and Smirnov by taking the 
classical limit of the construction in quantum integrable models
assuming a certain conjecture.
We propose another ${\cal D}$-free resolution of $A$ by extending
the construction in the classical finite dimensional 
integrable system associated with a certain family 
of hyperelliptic curves to infinite dimension 
assuming a similar conjecture. 
The relation of two constructions is given.
\end{abstract}

\section{Introduction}
In \cite{BBS} Babelon, Bernard and Smirnov (BBS), by considering the classical 
limit of a model in two dimensional integrable quantum field theory,
have studied the space of KdV fields $A={\mathbb C}[u,u',...]$ as a module
over the ring of commuting derivations 
${\cal D}={\mathbb C}[\partial_1,\partial_3,...]$, where $\partial_i$ acts
on $u$ according as the KdV-hierarchy (see section 2):
\bea
&&
\partial_i u=S'_{i+1}(u),
\quad
S_{i+1}(u)\in A.
\non
\ena
Assuming the conjecture that $A$ is generated over ${\cal D}$ by 
${\mathbb C}[S_2,S_4,...]$ they have constructed
 a ${\cal D}$-free resolution of $A$.
In particular all ${\cal D}$-linear relations among 
monomials of $\{S_{2i}\}$ are determined. 
They are called null vectors in \cite{BBS}. For example
the first two non-trivial null vectors are 
\bea
&&
\partial_3 S_2-\partial_1 S_4=0,
\quad
\partial_1^2 S_2-4S_4+6S_2^2=0,
\label{null-2}
\ena
which give the KdV equation for $S_2$:
\bea
&&
\partial_3 S_2=\frac{1}{4}\partial_1^3 S_2+3S_2\partial_1(S_2).
\non
\ena

In \cite{NS1} the affine ring $A_g$ of the affine Jacobian 
of a hyperelliptic curve of genus $g$ is studied as a module 
over the ring 
${\cal D}_g={\mathbb C}[\partial_1,\partial_3,...,\partial_{2g-1}]$
 of invariant vector fields on the Jacobian.
Assuming some conjecture a ${\cal D}_g$-free resolution
of $A_g$ has been constructed. 
Although the conjecture is verified only for $g\leq 3$ \cite{N3} up to now,
 this construction exhibits a remarkable consistency with other results.
For example it recovers the character of $A_g$ \cite{NS1} 
and the cohomologies of the affine Jacobian \cite{N1}. 
Since the $g\rightarrow \infty$ limit of 
$A_g$ is identified with $A$, in the present paper we directly construct a 
${\cal D}$-free resolution of $A$ extending
the construction for $A_g$.
Let $\tau(t)$ be the tau function of the KdV-hierarchy and
$\zeta_{i_1...i_n}=\partial_{i_1}\cdots\partial_{i_n}\log\,\tau(t)$,
$\partial_i=\partial/\partial t_{i}$.
Then the generators of $A$ over ${\cal D}$ of this construction
are given by the set of functions 
\bea
&&
1,\quad (i_1...i_n;j_1...j_n):=\det(\zeta_{i_k,j_l})_{1\leq k,l\leq n},
\quad
n\geq 1,
\non
\ena
where $i_1<\cdots<i_n$, $j_1<\cdots<j_n$.
The ${\cal D}$-linear relations among them are, for example,
\bea
&&
\partial_{j_1}(i_1i_2:j_2j_3)-\partial_{j_2}(i_1i_2:j_1j_3)+
\partial_{j_3}(i_1i_2:j_1j_2)=0.
\non
\ena
In a sense these are trivial relations since they hold if
$\tau(t)$ is replaced by an arbitrary function of $t_1,t_3,...$.
Notice that $S_{2n}=\partial_1\partial_{2n-1}$ and the null vectors
of BBS implies the bilinear form of the KdV equation for $\tau(t)$.
Thus two free resolutions of $A$ give quite different generators
and relations. However we prove that two constructions are equivalent
by showing the equivalence of two conjectures. 

The construction of \cite{BBS} is related with the quantum groups
at root of unity \cite{N2,FJKMMT}. While the construction of the present
paper is directly related with the geometry of Jacobian varieties and
the analysis of abelian functions \cite{N3,N4}. 
Thus the present result opens the way
to study the latter subjects in terms of the representation theory.
It is a quite interesting problem to extend conjectures and results of
\cite{NS1,N3} to the case of more general algebraic curves than that of
hyperelliptic curves based on this view point. Toward this direction
the results of \cite{NS2,SZ,DM} are important.

The present paper is organized in the following manner. 
In section 2 the space of KdV fields is defined. After reviewing
the boson-fermion correspondence in section 3, the construction of
the free resolution of $A$ due to Babelon, Bernard and Smirnov
is reviewed in section 4. In section 5 another construction of
the free resolution of $A$ is given. The relation of two
constructions is given in section 6. Finally in section 7 concluding
remarks are given.

\section{The Space of KdV Fields}
Let $A$ denote the differential algebra
\bea
&&
A=\mathbb{C}[u,u',u'',\cdots]
\ena
generated by $u=u^{(0)}$, $u'=u^{(1)}$, $u''=u^{(2)}$,  ... 
such that the derivation ${}^\prime$ acts as $(u^{(m)})'=u^{(m+1)}$
for any $m$.
The KdV hierarchy is the infinite number of compatible differential equations
given by
\bea
&&
\frac{\partial u}{\partial t_n}=S_{n+1}'(u),
\quad
n=1,3,5,...,
\ena
where $S_n(u)$ is the element of $A$ without constant term satisfying
the equation
\bea
&&
S_{n+2}'(u)=\frac{1}{4}S_{n}'''(u)-uS_{n}'(u)-\frac{1}{2}u'S_{n},
\quad
S_2(u)=-\frac{1}{2}u.
\ena
In particular ${}^\prime$ is identified with $\partial/\partial t_1$.
The KdV hierarchy defines the action of the commuting derivations 
$\partial_n$ on $A$ by
\bea
&&
\partial_n(u^{(k)})=S_{n+1}^{(k+1)}(u),
\label{action}
\ena
Thus $A$ is a ${\cal D}$-module, where 
${\cal D}=\mathbb{C}[\partial_1,\partial_3,\cdots]$.

\section{Free Fermions and Fock Spaces}
Let
$\psi_n$, $\psi_n^\ast$, $n\in 2\mathbb{Z}+1$, satisfy the 
relations 
\bea
&&
[\psi_m,\psi_n]_{+}=[\psi_m^\ast,\psi_n^\ast]_{+}=0,
\quad
[\psi_m^\ast,\psi_n]_{+}=\delta_{m,n},
\label{CACR}
\ena
where $[X,Y]_{+}=XY+YX$.
The vacuums $<m|$, $|m>$, $m\in 2\mathbb{Z}+1$ are defined by the conditions
\bea
&&
<m|\psi_n=0 \quad \text{for $n\leq m$},
\qquad
<m|\psi_n^\ast=0 \quad\text{for $n> m$},
\non
\\
&&
\psi_n|m>=0 \quad\text{for $n> m$},
\qquad
\psi_n^\ast|m>=0 \quad\text{for $n\leq m$}.
\non
\ena
They are related by
\bea
&&
\psi_m^\ast|m-2>=|m>,
\quad
<m-2|\psi_m=<m|.
\non
\ena
The Fock spaces $H_m$, $H_m^\ast$
are constructed from $|m>$ and $<m|$ respectively by the equal number of 
$\psi_k$ and $\psi^\ast_l$. The pairing between $H_m$ and $H_m^\ast$ are
defined by normalizing
\bea
&&
<m|m>=1.
\non
\ena
Let us set
\bea
&&
h_{-2k}=\sum_{n\in 2\mathbb{Z}+1}\psi_n\psi^\ast_{n+2k},
\non
\\
&&
T=\exp(-\sum_{k=1}^\infty \frac{1}{k}J_{2k}h_{-2k}),
\non
\ena
where $J_{2k}$ are commutative variables.
Notice that $<m|T=<m|$ for any $m$.
The boson-fermion correspondence gives the isomorphism of bosonic and fermionic
Fock spaces \cite{DJKM}
\bea
&&
H^\ast_{2m-1}\simeq \mathbb{C}[J_2,J_4,\ldots],
\label{isom1}
\\
&&
<2m-1|a \mapsto <2m-1|aT|2m-1>.
\non
\ena

\section{Babelon-Bernard-Smirnov's Construction}
In this section we review the results of \cite{BBS}. Their discovery
is that the fermionic description of the bosonic map $\ev_1$ defined below
(\ref{ev-1}) greatly simplifies the situation. Later it is understood that
such structure is intimately related with quantum groups at a root of unity 
\cite{N2,FJKMMT}.

Let
\bea
&&
\psi(z)=\sum_{n\in 2\mathbb{Z}+1}\psi_n z^{-n},
\quad
\psi^\ast(z)=\sum_{n\in 2\mathbb{Z}+1}\psi_n^\ast z^{n}.
\non
\ena
Define two operators $Q$ and $C$ by
\bea
&&
Q=\int\frac{dz}{2\pi i}\nabla(z)\psi(z),
\non
\\
&&
C=\int\frac{dz}{2\pi i}\psi(z)\frac{d\psi(z)}{dz}+
\int\int_{|z_2|>|z_1|}\frac{dz_1}{2\pi i}\frac{dz_2}{2\pi i}
\log\Big(1-(\frac{z_1}{z_2})^2\Big) \nabla(z_1)\nabla(z_2)\psi(z_1)\psi(z_2).
\non
\ena
Here the simple integral signifies to take the coefficient of $z^{-1}$ and
the double integral signifies to take that of $(z_1z_2)^{-1}$ when the integrand@is expanded at the region $|z_1/z_2|<1$.
These operators are maps of the following spaces
\bea
&&
Q:\, {\cal D}\otimes H^\ast_n \lar {\cal D}\otimes H^\ast_{n+2},
\non
\\
&&
C:\, {\cal D}\otimes H^\ast_n \lar {\cal D}\otimes H^\ast_{n+4}.
\non
\ena
They satisfy
\bea
&&
[Q,C]=0,
\quad
Q^2=0.
\label{QCrelation}
\ena
Let us introduce new variables $\bS$ by
\bea
&&
\exp(-\sum_{k=1}^\infty \frac{1}{k}J_{2k}z^{-2k})=
\sum_{n=0}^\infty \bS_{2n}z^{-2n}.
\non
\ena
By specifying the degrees as $\deg\, J_{2k}=2k$, 
$\bS_{2n}$ is a homogeneous polynomial of $J_{2k}$'s of 
degree $2k$ and has the form
\bea
&&
\bS_{2n}=\frac{-J_{2n}}{2n}+\cdots,
\non
\ena
where $\cdots$ part does not contain $J_{2n}$. In particular we have
the isomorphism of polynomial rings
\bea
&&
\mathbb{C}[J_2,J_4,\ldots]\simeq \mathbb{C}[\bS_2,\bS_4,\ldots].
\label{isom2}
\ena
The composition of (\ref{isom1}) and (\ref{isom2}) gives the isomorphism
\bea
&&
H_{-1}^\ast \simeq \mathbb{C}[\bS_2,\bS_4,\ldots].
\non
\ena
We identify these two spaces by this isomorphism.
We define  a map
\bea
&&
\ev_1:{\cal D}\otimes \mathbb{C}[\bS_2,\bS_4,\ldots]\lar A
\label{ev-1}
\ena
by
\bea
&&
P(\partial)\otimes \bS_2^{\alpha_2}\bS_4^{\alpha_4}\cdots
\mapsto
P(\partial)(S_2^{\alpha_2}S_4^{\alpha_4}\cdots).
\non
\ena

Then Babelon, Bernard and Smirnov have proved

\begin{theorem}\text{(}\cite{BBS,BBT}\text{)}\label{ker-ev1}
\bea
&& 
Q({\cal D}\otimes H_{-3}^\ast)+C({\cal D}\otimes H_{-5}^\ast)
\subset \Ker\, \ev_1.
\non
\ena
\end{theorem}

They conjectured

\begin{conj}\label{conj1}
The map $\ev_1$ is surjective.
\end{conj}

We set 
\bea
&&
B=\frac{{\cal D}\otimes H_{-1}^\ast}{Q({\cal D}\otimes H_{-3}^\ast)+C({\cal D}\otimes H_{-5}^\ast)}.
\non
\ena
Since $Q^2=0$ we have the complex
\bea
&&
\cdots
\stackrel{Q}{\lar}
{\cal D}\otimes H_{-5}^\ast 
\stackrel{Q}{\lar}
{\cal D}\otimes H_{-3}^\ast
\stackrel{Q}{\lar}
{\cal D}\otimes H_{-1}^\ast
\lar
0.
\label{seq1}
\ena
Since $C$ and $Q$ commute, it induces the following complex;
\bea
&&
\cdots
\stackrel{Q}{\lar}
\frac{{\cal D}\otimes H_{-5}^\ast}{C({\cal D}\otimes H_{-9}^\ast)} 
\stackrel{Q}{\lar}
\frac{{\cal D}\otimes H_{-3}^\ast}{C({\cal D}\otimes H_{-7}^\ast)} 
\stackrel{Q}{\lar}
\frac{{\cal D}\otimes H_{-1}^\ast}{C({\cal D}\otimes H_{-5}^\ast)} 
\lar
0,
\label{seq2}
\ena
because $[Q,C]=0$.
Finally we have the complex
\bea
&&
\cdots
\stackrel{Q}{\lar}
\frac{{\cal D}\otimes H_{-5}^\ast}{C({\cal D}\otimes H_{-9}^\ast)} 
\stackrel{Q}{\lar}
\frac{{\cal D}\otimes H_{-3}^\ast}{C({\cal D}\otimes H_{-7}^\ast)} 
\stackrel{Q}{\lar}
\frac{{\cal D}\otimes H_{-1}^\ast}{C({\cal D}\otimes H_{-5}^\ast)} 
\lar
B
\lar
0,
\label{seq3}
\ena
where the map to $B$ is the natural projection.

\begin{prop}
(i) For $n\geq 0$
$$
\frac{{\cal D}\otimes H_{-2n-1}^\ast}{C({\cal D}\otimes H_{-2n-5}^\ast)}
$$
is a free ${\cal D}$-module.

\noindent
(ii) The complex (\ref{seq3}) is exact.
\end{prop}

By this proposition (\ref{seq3}) gives a ${\cal D}$-free resolution of $B$.

The statement (i) of this proposition is proved in \cite{BBT}. 
We recall the change of fermions used there for further use.
In the component form $Q$ and $C$ are written as
\bea
&&
Q=\sum_{n=1}^\infty \partial_{2n-1}\psi_{-(2n-1)},
\non
\\
&&
C=\sum_{n=1}^\infty \Big(2(2n-1)\psi_{2n-1}-
\sum_{l=1}^\infty P_{n,l}(\partial)\psi_{2n-1-l}\Big)
\psi_{-(2n-1)},
\non
\ena
where we set
\bea
&&
P_{n,l}(\partial)=\sum_{i+j=l+1, j<n}
\frac{1}{n-j}\partial_{2i-1}\partial_{2j-1}.
\non
\ena
We define
\bea
&&
\tpsi_{-(2n-1)}=\psi_{-(2n-1)}
\quad\text{for $n\geq 1$},
\non
\\
&&
\tpsi_{2n-1}=2(2n-1)\psi_{2n-1}-\sum_{l=1}^\infty P_{n,l}(\partial)\psi_{2n-1-2l}
\quad\text{for $n\geq 1$},
\non
\ena
Write these relations as
\bea
&&
\tpsi_i=\sum_{j\in 2\mathbb{Z}+1} d_{ij}\psi_j,
\non
\ena
and set $D=(d_{ij})$ which is an invertible triangular matrix. Set
\bea
&&
D'=(d'_{ij})={}^t(D^{-1}),
\non
\\
&&
\tpsi^\ast_i=\sum_{j}d'_{ij}\psi_{j}^\ast.
\non
\ena
Then $\{\tpsi_i,\tpsi_j^\ast\}$ satisfy the canonical anti-commutation 
relations (\ref{CACR}).
 Moreover the vacuums $<m|$, $|m>$ for $\{\psi_i,\psi_j^\ast\}$ 
become the vacuums for $\{\tpsi_i,\tpsi_j^\ast\}$. 
We denote the Fock spaces of 
$\{\tpsi_i,\tpsi_j^\ast\}$ by $\tH_m$, $\tH_m^\ast$. Then
\bea
&&
Q=\sum_{n=1}^\infty \partial_{2n-1}\tpsi_{-(2n-1)},
\quad
C=\sum_{n=1}^\infty \tpsi_{2n-1}\tpsi_{-(2n-1)},
\non
\ena
and we have isomorphisms
\bea
&&
{\cal D}\otimes H_{-2n-1}^\ast \simeq {\cal D}\otimes \tH_{-2n-1}^\ast,
\non
\\
&&
\frac{{\cal D}\otimes H_{-2n-1}^\ast}{C({\cal D}\otimes H_{-2n-5}^\ast)}
\simeq
{\cal D}\otimes \frac{\tH_{-2n-1}^\ast}{C\tH_{-2n-5}^\ast},
\non
\ena
for any integer $n$.
The statement (ii) follows from the following lemmas in a similar manner
 to Theorem 4.3 of \cite{N2}. The lemmas can also be proved similarly 
to Lemma 4.4 and 4.5 of \cite{N2}.

\begin{lemma}
The complex (\ref{seq1}) is exact at ${\cal D}\otimes H_{-2n-1}$, $n\geq 1$.
\end{lemma}

\begin{lemma}
The map
\bea
&&
C:\, \tH_{-2m-1}\lar \tH_{-2m+3}
\non
\ena
is injective for $m\geq 0$.
\end{lemma}

\begin{cor}
If we assume Conjecture \ref{conj1}, then $A\simeq B$ and
(\ref{seq3}) gives a ${\cal D}$-free resolution of $A$,
\bea
&&
\cdots
\stackrel{Q}{\lar}
{\cal D}\otimes \frac{\tH_{-5}^\ast}{C\tH_{-9}^\ast} 
\stackrel{Q}{\lar}
{\cal D}\otimes \frac{\tH_{-3}^\ast}{C\tH_{-7}^\ast} 
\stackrel{Q}{\lar}
{\cal D}\otimes \frac{\tH_{-1}^\ast}{C\tH_{-5}^\ast} 
\stackrel{\ev_1}{\lar}
A
\lar
0.
\label{resol1}
\ena
\end{cor}
\vskip2mm

\noindent
{\it Proof.} 
Conjecture \ref{conj1} implies that the map $B\rightarrow A$ induced
from $\ev_1$ is surjective.
Then the injectivity follows by comparing characters.
For a graded vector space $V=\oplus V_n$ with $\dim\,V_n<\infty$
 we define the character of $V$ by
\bea
&&
\ch\,V=\sum q^n \dim\,V_n.
\non
\ena
For $H_n^\ast$ and $\tilde{H}_n^\ast$ we assign
\bea
\deg\,\psi_n=n,
\quad
\deg\,\psi^\ast_n=-n,
\quad
\deg\,<2m-1|=m^2,
\quad
\deg\,\partial_i=i.
\non
\ena
Then 
\bea
&&
\ch\,H^\ast_{-2m+1}=\ch\,\tilde{H}^\ast_{-2m+1}=
\frac{q^{m^2}}{\prod_{i=1}^\infty(1-q^{2i})}.
\non
\ena
For $A$ we define $\deg\,u^{(i)}=2+i$. Then 
\bea
&&
\ch\,A=\frac{1-q}{\prod_{i=1}^\infty(1-q^i)}
\non
\ena
and the map $\ev_1$ preserves grading.
Using the free resolution (\ref{seq3}) of $B$ we have
\bea
&&
\ch\, B=\frac{1-q}{\prod_{i=1}^\infty(1-q^i)}=\ch\, A,
\non
\ena
which completes the proof.
\qed

\section{Another Construction of Free Resolution}
In this section we shall generalize the construction of \cite{NS1}
to the case of infinite degrees of freedom.

Let $\tau(t)=\tau(t_1,t_3,\ldots)$ be the tau function of the KdV-hierarchy 
\cite{DJKM}.
We set
\bea
&&
\zeta_i=\partial_i \log\, \tau(t),
\quad\zeta_{ij}=\partial_i\partial_j \log\, \tau(t),
\quad
\partial_i=\frac{\partial}{\partial t_i}.
\non
\ena
Notice that $\zeta_{ij}$ can be expressed as a differential polynomial
of $u=\zeta_{11}$ and thereby is contained in $A$.
Then
\bea
&&
d\zeta_{i}=
\sum_{j:odd} \zeta_{ij} dt_{j}
\non
\ena
is a $1$-form with the coefficients in $A$.

Let
\bea
&&
\alpha_{2n-1}=\tpsi_{-(2n-1)},
\quad
\beta_{2n-1}=\tpsi_{2n-1},
\quad
n\geq 1.
\non
\ena
Then
\bea
&&
Q=\sum_{n=1}^\infty \partial_{2n-1}\alpha_{2n-1},
\quad
C=\sum_{n=1}^\infty \beta_{2n-1}\alpha_{2n-1}.
\non
\ena
For $N\geq 1$ set
\bea
&&
\tH^\ast_{-1}(N)=\sum_{k=0}^N\sum 
{\mathbb C}<-2N-1|\alpha_{i_{N-k}} \cdots \alpha_{i_1}\beta_{j_k} \cdots \beta_{j_1},
\non
\ena
where the second summation is over all odd integers satisfying
\bea
&&
2N+1>i_{N-k}>\cdots>i_1\geq 1,
\quad
j_k>\cdots>j_1\geq 1.
\non
\ena
Set $H^\ast_{-1}(0)={\mathbb C}<-1|$.
We use the notation like 
\bea
&&
\alpha_I=\alpha_{i_{N-k}} \cdots \alpha_{i_1},
\non
\ena
for $I=(i_{N-k},\ldots,i_1)$.

For $N<N'$ we have the inclusion
\bea
&&
\tH_{-1}^\ast(N) \subset \tH_{-1}^\ast(N'),
\label{inclusion}
\\
&&
<-2N-1|a
\mapsto
<-2N'-1|\alpha_{2N'-1}\alpha_{2N'-3}\cdots \alpha_{2N+1}a.
\non
\ena
Thus $\{\tH_{-1}^\ast(N)\}$ defines an increasing filtration of $H^\ast_{-1}$:
\bea
&&
\tH_{-1}^\ast =\cup_{N=0}^\infty \tH_{-1}^\ast(N).
\non
\ena
We define a map of ${\cal D}$-modules
\bea
&&
\ev_2:\, {\cal D} \otimes \tH_{-1}^\ast \lar A,
\non
\ena
as follows.

Let 
\bea
&&
\vol=\cdots\wedge dt_5\wedge dt_3\wedge dt_1,
\quad
\Omega^{\frac{\infty}{2}}=\mathbb{C}\,\vol,
\non
\ena
and $\Omega^{\frac{\infty}{2}-p}$ be the vector space generated by differential
forms which are obtained from $\vol$ by removing $p$ $dt_i$'s.
We define the action of ${\cal D}$ on $A\otimes \Omega^{\frac{\infty}{2}}$
by
\bea
&&
P(F\vol)=P(F)\vol,
\quad
P\in {\cal D},
\quad
F\in A,
\non
\ena
where we omit the tensor symbol for simplicity.
Then we have the isomorphism of ${\cal D}$-modules
\bea
&&
A\otimes \Omega^{\frac{\infty}{2}} \simeq A,
\non
\\
&&
F\vol \mapsto F.
\ena

Let $v\in \tH_{-1}^\ast(N)$ be of the form
\bea
&&
v=<-2N-1|\,\alpha_I\beta_J,
\quad
I=(i_{N-k},\ldots,i_1),
\quad
J=(j_k,\ldots,j_1),
\non
\ena
and $P\in {\cal D}$.
We define
\bea
&&
\ev_2(P \otimes v)\,\vol=
P(\cdots\wedge dt_{2N+3}\wedge dt_{2N+1}
\wedge dt_I \wedge d\zeta_J),
\non
\ena
where $dt_I=dt_{i_{N-K}}\wedge\cdots\wedge dt_{i_1}$ etc. 
We can write $\ev_2$ more explicitly using certain determinants.  
Write
\bea
&&
\cdots\wedge dt_{2N+3}\wedge dt_{2N+1}
\wedge dt_I \wedge d\zeta_J=F_{IJ}\,\vol,
\quad
F_{IJ}\in A.
\non
\ena
Let 
\bea
&&
I^{c}=\{1,3,\ldots,2N-1\}\backslash I
=\{l_k>\cdots>l_1\}
\non
\ena
and $\sgn(I,I^{c})$ be the sign of the permutation 
\bea
&&
(2N-1,\ldots,3,1)\lar (I,I^{c}).
\non
\ena
Then 
\bea
&&
\ev_2(P \otimes v)=P(F_{IJ}),
\non
\\
&&
F_{IJ}=\sgn(I,I^{c})
\det(\zeta_{l_aj_b})_{1\leq a,b\leq k}.
\non
\ena
One can immediately check, using (\ref{inclusion}),
 that the definition of $\ev_2$ does not depend on the choice of $N$ such that
$v\in \tH_{-1}^\ast(N)$.

\begin{prop}\label{ker-ev2}
(i) $\ev_2\Big(Q({\cal D}\otimes \tH_{-3}^\ast)\Big)=0$.

\noindent
(ii) $\ev_2\Big(C({\cal D}\otimes \tH_{-5}^\ast)\Big)=0$.
\end{prop}
\vskip2mm

\noindent
{\it Proof.} (i)
We define
\bea
&&
d:\, 
A \otimes \Omega^{\frac{\infty}{2}-p}
\lar 
A \otimes \Omega^{\frac{\infty}{2}-p+1},
\non
\ena
by
\bea
&&
d(F\otimes w)=
\sum_{n=1}^\infty \partial_{2n-1} F \otimes w \wedge dt_{2n-1},
\quad
F\in A,
\quad
w\in \Omega^{\frac{\infty}{2}-p}.
\non
\ena
Then 
\bea
&&
d^2=0,
\quad
d\, \Omega^{\frac{\infty}{2}-p}=0,
\non
\\
&&
d(w_1\wedge w_2)=w_1\wedge dw_2+(-1)^q dw_1 \wedge w_2,
\quad
w_1\in A \otimes \Omega^{\frac{\infty}{2}-p-q},
\quad
w_2\in A \otimes \Omega^{q},
\non
\ena
where $\Omega^{q}$ is the space of $q$-forms of $dt_1$, $dt_3$,... and $dw_2$ is defined in an obvious manner.

For $I=(i_k,\ldots,i_1)$ we set $|I|=k$.
Let
\bea
&&
v=<-2N-1|\, \alpha_I\beta_J\in \tH_{-3}^\ast,
\quad
|I|+|J|=N-1,
\non
\ena
and $P\in {\cal D}$.
Then
\bea
&&
Q(P \otimes v)=
\sum_{n=1}^\infty \partial_{2n-1}P 
\otimes 
<-2N-1|\, \alpha_I\beta_J\alpha_{2n-1},
\non
\ena
and
\bea
\ev_2\Big(Q(P \otimes v)\Big)\vol
&=&
\sum_{n=1}^\infty 
\partial_{2n-1}P(\cdots\wedge dt_{2N+1} \wedge 
dt_I \wedge d\zeta_J \wedge dt_{2n-1})
\non
\\
&=&
P\Big(d(\cdots\wedge dt_{2N+1} \wedge 
dt_I \wedge d\zeta_J)\Big)
\non
\\
&=&
P\Big((-1)^{N-1}d(\cdots\wedge dt_{2N+1}) \wedge 
dt_I \wedge d\zeta_J
+\cdots\wedge dt_{2N+1} \wedge d(dt_I \wedge d\zeta_J)
\Big)
\non
\\
&=&
0.
\non
\ena

\noindent
(ii) Let
\bea
&&
v=P\otimes <-2N-1|\, \alpha_I\beta_J\in {\cal D} \otimes \tH_{-5}^\ast.
\non
\ena
Then
\bea
\ev_2(Cv)\vol
&=&
\ev_2(-P\otimes \sum_{n=1}^\infty 
<-2N-1|\, \alpha_I\beta_J\alpha_{2n-1}\beta_{2n-1})
\non
\\
&=&
-P(\cdots\wedge dt_{2N+1} \wedge 
dt_I \wedge d\zeta_J \wedge \sum_{n=1}^\infty dt_{2n-1}\wedge d\zeta_{2n-1})
\non
\\
&=&
0,
\non
\ena
since
\bea
&&
\sum_{n=1}^\infty dt_{2n-1}\wedge d\zeta_{2n-1}
=\sum_{n,m=1}^\infty (\partial_{2m-1}\partial_{2n-1}\log\, \tau)
dt_{2n-1}\wedge dt_{2n-1}
=0.
\non
\ena
\qed

By (ii) of Proposition \ref{ker-ev2} we have a map of ${\cal D}$-modules
\bea
&&
{\cal D}\otimes \frac{\tH_{-1}^\ast}{C\tH_{-5}^\ast} \stackrel{ev_2}{\lar} A.
\label{surj-ev2}
\ena

Notice that the proof of Proposition \ref{ker-ev2} is much simpler than 
 that of Theorem \ref{ker-ev1} in \cite{BBS,BBT}.

The following theorem is proved in the next section.

\begin{theorem}\label{resol2}
If we assume Conjecture \ref{conj1} then the map (\ref{surj-ev2})
is surjective.  In particular if we replace $\ev_1$ by $\ev_2$ in 
(\ref{resol1}) then it gives a ${\cal D}$-free resolution of $A$.
\end{theorem}

\section{Relation of Two Constructions}
In this section we show

\begin{theorem}\label{equivalence}
The map $\ev_1$ is surjective if and only if the map $\ev_2$ is surjective.
\end{theorem}

The remaining part of this section is devoted to the proof of this theorem.
Let us set
\bea
&&
\omega_{n,m}=\ev_1({\bar \omega}_{n,m}),
\quad
{\bar \omega}_{n,m}=<-1|\psi_{n}\psi_{-m}^\ast T|-1>,
\quad
m,n\geq 1,
\non
\\
&&
\omega_{n}=\sum_{m:odd} \omega_{n,m}dt_{m}.
\non
\ena

\begin{lemma}
The $1$-form $\omega_{n}$ is closed.
\end{lemma}
\vskip2mm

\noindent
{\it Proof.}
By Theorem \ref{ker-ev1}
\bea
&&
\ev_1\left(<-1|\psi_{n}\psi_{-m_1}^\ast\psi_{-m_2}^\ast QT|-1>\right)=0.
\label{Q=0}
\ena
We substitute
\bea
&&
Q=\sum_{i:odd} \partial_{i}\psi_{-i}
\non
\ena
into (\ref{Q=0}) and get
\bea
&&
\partial_{m_2}\omega_{n,m_1}
-\partial_{m_1}\omega_{n,m_2}=0
\non
\ena
which proves the lemma.
\qed

By the lemma $\omega_{n}$ should be written as $d\eta_{n}$ for 
some function $\eta_{n}$ which is not necessarily in $A$.
We shall find the explicit form of $\eta_{n}$
and study its properties. 

Let 
\bea
&&
\Psi(z)=\frac{\tau(t-[z^{-1}])}{\tau(t)}\e^{\xi(t,z)},
\quad
\xi(t,z)=\sum_{n=1}^\infty t_{2n-1}z^{2n-1},
\non
\ena
be the wave function of the KdV-hierarchy \cite{DJKM,BBT},
where $[z^{-1}]=(z^{-1},\frac{z^{-3}}{3},\frac{z^{-5}}{5},\ldots)$.
Set
\bea
&&
S(z)=\sum_{n=0}^\infty S_{2n}z^{-2n},
\quad
S_0=1.
\non
\ena
Notice that $S_{2n}=\partial_1\partial_{2n-1}\log\,\tau(t)$.
Then, by the bilinear identity for $\tau(t)$ \cite{DJKM}, we have 
(see Remark 1, p391, in \cite{BBT} for example)
\bea
&&
S(z)=\frac{\tau(t-[z^{-1}])\tau(t+[z^{-1}])}{\tau(t)^2}.
\label{S-tau}
\ena
Define $X(z)$ by (\cite{BBT} p388)
\bea
&&
X(z)=\frac{-1}{2}\log\, S(z)+\log\, \Psi(z).
\non
\ena
Using (\ref{S-tau}) we have
\bea
X(z)=\frac{1}{2}\Big(\log\, \tau(t-[z^{-1}])-\log\, \tau(t+[z^{-1}])\Big)
+\xi(t,z).
\label{X-tau}
\ena
Let us set
\bea
&&
\eta(z)=z^{-1}\Big(-X(z)+\xi(t,z)\Big).
\non
\ena
By (\ref{X-tau}) $\eta(z)$ is expanded into negative even powers of $z$,
\bea
&&
\eta(z)=\sum_{n=1}^\infty \eta_{2n-1} z^{-2n}.
\non
\ena
Then

\begin{prop}
(i) $d\eta_{2n-1}=\omega_{2n-1}$.

\noindent
(ii) We have
\bea
&&
\eta_{2n-1}=\frac{1}{2n-1}\zeta_{2n-1}+a_{2n-1},
\label{descript-eta}
\ena
for some $a_{2n-1}\in A$.
\end{prop}
\vskip2mm

\noindent
{\it Proof.}
We use the following lemma (\cite{BBT} p392 Remark 3).

\begin{lemma} Let $\nabla(w)=\sum_{n=1}^\infty \partial_{2n-1} w^{-2n}$.
 Then
\bea
&&
\nabla(w) X(z)=\frac{z}{w^2-z^2}\frac{S(w)}{S(z)},
\quad
|w|>|z|.
\non
\ena
\end{lemma}

By this lemma we get
\bea
&&
\nabla(w) \eta(z)=\frac{1}{z^2-w^2}\Big(\frac{S(w)}{S(z)}-1\Big),
\quad
|w|>|z|.
\label{nabla-eta}
\ena
Since the right hand side of this equation is regular at $z^2=w^2$, 
(\ref{nabla-eta}) is valid at $|z|>|w|$.
Let ${\bar S}(z)=\sum_{n=0}^\infty {\bar S}_{2n}z^{-2n}$ with ${\bar S}_0=1$.
Using
\bea
&&
T\psi(z)T^{-1}={\bar S}(z)\psi(z),
\quad
T\psi^\ast(z)T^{-1}={\bar S}(z)^{-1}\psi^\ast(z),
\non
\\
&&
<-1|\psi(z)\psi^\ast(w)|-1>=\frac{zw}{z^2-w^2},
\quad
|z|>|w|,
\non
\ena
we have
\bea
&&
<-1|\psi(z)\psi^\ast(w)T|-1>=\frac{{\bar S}(w)}{{\bar S}(z)}\frac{zw}{z^2-w^2},
\quad
|z|>|w|.
\non
\ena
On the other hand, expanding in $z$ and $w$, we have
\bea
&&
<-1|\psi(z)\psi^\ast(w)T|-1>
=
\sum_{m,n=1}^\infty
{\bar \omega}_{2n-1,2m-1}z^{-2n+1}w^{-2m+1}
+
\frac{zw}{z^2-w^2},
\quad
|z|>|w|.
\non
\ena
Thus
\bea
\frac{1}{z^2-w^2}\Big(\frac{{\bar S}(w)}{{\bar S}(z)}-1\Big)
&=&
(zw)^{-1}<-1|\psi(z)\psi^\ast(w)T|-1>-\frac{1}{z^2-w^2}
\non
\\
&=&
\sum_{m,n=1}^\infty
{\bar \omega}_{2n-1,2m-1}z^{-2n}w^{-2m}.
\non
\ena
Taking $\ev_1$ we have
\bea
&&
\partial_{2m-1}\eta_{2n-1}=\omega_{2n-1,2m-1}.
\non
\ena
\qed
\vskip3mm

\noindent
{\it Proof of Theorem \ref{equivalence}}
\par

We denote by $\vol(i_1,\ldots,i_k)$
the differential form which is obtained from $\vol$ by removing
$dt_{i_1}$,...,$dt_{i_k}$.

Suppose that $\ev_1$ is surjective. 
Notice that
\bea
&&
\ev_1:\, {\cal D}\otimes H_{-1}^\ast \lar A,
\non
\ena
is given by
\bea
&&
\ev_1\Big(P\otimes 
<-1|\psi_{2i_1-1} \cdots \psi_{2i_k-1}\psi_{2j_k-1}^\ast \cdots 
\psi_{2j_1-1}^\ast\Big)=
P\Big(\det(\omega_{2i_a-1,2j_b-1})_{1\leq a,b\leq k}\Big).
\non
\ena
Thus the space
\bea
&&
\Big(\frac{A}{\sum_{n=1}^\infty \partial_{2n-1}A}\Big)\,\vol
\label{mod-space}
\ena
is generated by all elements of the form
\bea
&&
\vol(i_1,\ldots,i_k)\wedge \omega_{2j_1-1}\wedge\cdots\wedge \omega_{2j_k-1},
\label{gen-1}
\ena
as a vector space over $\mathbb{C}$.
We substitute (\ref{descript-eta}) into (\ref{gen-1}). 
Then the space $(\ref{mod-space})$ is generated by the elements 
of the form
\bea
&&
\vol(i_1,\ldots,i_k)\wedge d\zeta_{2j_1-1}\wedge\cdots\wedge d\zeta_{2j_k-1},
\non
\ena
since the terms containing $da_{2r-1}$ belong to the denominator 
$(\sum_{n=1}^\infty \partial_{2n-1}A)\vol$.
Since (\ref{mod-space}) generates $A$ over ${\cal D}$, $\ev_2$ is surjective.
The converse is similarly proved.
\qed

As a corollary of Theorem \ref{equivalence}, Theorem \ref{resol2} in the
previous section is proved.

\section{Concluding Remarks}
From the view point of conformal field theories 
and their integrable deformations \cite{Z,EY,SY}
the space $A$ corresponds to the descendent fields of the identity operator.
It is possible to consider the spaces corresponding to descendents of 
other primary fields \cite{BBS}. It is an interesting problem to construct
their ${\cal D}$-free resolutions. In the quantum case free resolutions
are constructed for those spaces \cite{BBS,N2}. 
In particular the spaces become free modules in "odd cases".
The classical and even the finite dimensional cases 
are expected to have similar structures. 
Geometrically, to consider non-identity primary fields 
corresponds to consider the spaces of sections of certain 
non-trivial flat line bundles over affine Jacobians in stead 
of affine rings \cite{NS1}.

\end{document}